\documentclass[aps,prd,twocolumn,superscriptaddress,showpacs]{revtex4}
\usepackage{epsfig,epsf}
\usepackage{amsmath}
\usepackage{amsthm}
\usepackage{amsfonts}
\usepackage{amssymb}
\usepackage{dsfont}
\usepackage{multirow}
\usepackage{appendix}
\usepackage{slashed}
\usepackage[active]{srcltx}
\usepackage{psfrag}

\begin{document}

\title{ Exploring $X(5568)$ as a meson molecule}
\date{\today}
\author{S.~S.~Agaev}
\affiliation{Department of Physics, Kocaeli University, 41380 Izmit, Turkey}
\affiliation{Institute for Physical Problems, Baku State University, Az--1148 Baku,
Azerbaijan}
\author{K.~Azizi}
\affiliation{Department of Physics, Do\v{g}u\c{s} University, Acibadem-Kadik\"{o}y, 34722
Istanbul, Turkey}
\author{H.~Sundu}
\affiliation{Department of Physics, Kocaeli University, 41380 Izmit, Turkey}

\begin{abstract}
The parameters, i.e. the mass and current coupling of the exotic $X(5568)$
state observed by the D0 Collaboration as well as the decay width of the
process $X \to B_s^{0}\pi^{+}$ are explored using $B\overline{K}$ molecule
assumption on its structure. Employed computational methods include QCD
two-point and light-cone sum rules, latter being considered in the
soft-meson approximation. The obtained results are compared with the data of
the D0 Collaboration as well as with the predictions of the
diquark-antidiquark model. This comparison strengthens a diquark-antidiquark
picture for the $X(5568)$ state rather than a meson molecule structure.
\end{abstract}

\pacs{14.40.Rt, 12.39.Mk,  11.55.Hx}
\maketitle


\section{Introduction}

\label{Sec:Int}

Both the experimental and theoretical investigations of exotic hadronic
states, i.e. particles that are beyond the schemes of traditional hadron
spectroscopy, seem already formed as one of the interesting and rising
branches of the high energy physics. Since the pioneering discovery of the
charmoniumlike resonance $X(3872)$ by Belle Collaboration in 2003 \cite%
{Belle:2003} (see, also Refs. \cite{D0:2004,CDF:2004,Babar:2005}), numerous
exotic particles were observed and studied. Now they are organized as the $%
XYZ$ family of the exotic particles. During passed years, experimental and
theoretical investigation of these particles achieved evident successes in
measurements of their masses and decay widths, in exploring spins and
parities, as well as in creating various theoretical models and schemes to
reveal their internal structure and compute corresponding parameters (see
for instance \cite%
{Swanson:2006st,Klempt:2007cp,Godfrey:2008nc,Voloshin:2007dx,Nielsen:2010,
Faccini:2012pj,Esposito:2014rxa,Chen:2016} and references therein).

Recently, due to the observation of the resonance structure reported by the
D0 Collaboration in Ref.\ \cite{D0:2016mwd}, this interest to unusual
hadrons is renewed. Really, in accordance with this observation, the new
narrow resonance $X(5568)$ (in what follows denoted as $X_b(5568)$ or $X_b$)
contains four valence quarks and belongs to the class of the exotic states.
But what is important, the $X_b(5568)$ is probably the first hadronic state
built of four-quarks of different flavors, namely $b,\,s,\,u$ and $d$
quarks, which makes it a very interesting object for the experimental and
theoretical studies. The D0 Collaboration measured its mass, width of the
dominant decay channel $X_b \to B_s\pi$, and assigned the quantum numbers $%
J^{PC}=0^{++}$ to this resonance as preferable ones. Moreover, in Ref.\ \cite%
{D0:2016mwd} some suggestions concerning the structure of the $X_b(5568)$
were made, as well. Thus, the $X_b(5568)$ may be considered as a
diquark-antidiquark bound state within the tetraquark model of the exotic
states. Alternatively, $X_b$ may be studied as a molecule composed of $B$
and $\overline{K}$ mesons.

Nevertheless, despite the experimental information on the $X_b(5568)$ state
provided by the D0 Collaboration, the LHCb Collaboration has not still
confirmed the D0 observation, as they announced in a preliminary report \cite%
{LHCb:2016}. This situation demands intensive experimental and theoretical
explorations of its features to answer questions on the nature of the
observed state.

The information of the D0 Collaboration led to burst of activity in the
relevant theoretical studies. It started immediately after discovery of the
exotic $X_b(5568)$ state and employed almost all possible scenarios to
explain its observed parameters. In fact, in Refs.\ \cite%
{Agaev:2016mjb,Agaev:2016ijz} we calculated the mass, decay constant and
width of the $X_b(5568)$ state within the diquark-antidiquark picture $%
X_b=[su][\bar{b}\bar{d}]$ considering the exotic state with positive charge.
Our results for the mass $m_{X_b}$, and for the width of its decay $%
\Gamma(X_{b}\to B_{s}^{0}\pi^{+})$ are in a good agreement with the
experimental data.

In Ref.\ \cite{Agaev:2016lkl} we extended our investigation of the new
family of the four-quark exotic states by considering the charmed partner of
the $X_b$ state, and assuming that the $X_c$ state is composed of the $%
c,\,s,\, u$ and $d $ quarks. We also supposed that $X_c$ possesses the same
quantum numbers as $X_b$. We computed the mass, decay constant and width of
the decays $X_c \to D_{s}^{-}\pi^{+}$ and $X_c \to D^{0}K^{0}$ considering $%
X_c$ as $[su][\bar{c}\bar{d}]$ diquark-antidiquark state and employing two
forms for the interpolating currents.

The mass of the $X_b$ state was also estimated in Refs.\ \cite%
{Wang:2016tsi,Chen:2016mqt,Zanetti:2016wjn,Wang:2016mee} using
the diquark-antidiquark model. The width of the decay channel $%
X_{b}^{\pm}(5568) \to B_{s}\pi^{\pm}$ was calculated in Ref.\ \cite%
{Dias:2016dme,Wang:2016wkj} in the framework of the three-point QCD sum rule
approach, and good agreement between the theoretical predictions for $%
\Gamma(X_{b}^{\pm} \to B_{s}^{0}\pi^{\pm})$ and experimental data was
reported. As a bound state of the $B$ and $\overline{K}$ mesons the exotic $%
X_b$ particle was studied in Ref.\ \cite{Xiao:2016mho}. Some questions of
quark-antiquark organization of $X_b$ and its partners were addressed in
Ref.\ \cite{Liu:2016ogz}.

The controversial information of the D0 and LHCb collaborations triggered an
appearance of interesting theoretical works devoted to analysis of the $X_b$
physics. In these papers numerous aspects of the $X_b$ problems, including
investigations of its structure and parameters, analysis some of production
mechanisms are covered. For details and further explanations we refer to
original works \cite%
{He:2016yhd,Jin:2016cpv,Stancu:2016sfd,Burns:2016gvy,Tang:2016pcf,Guo:2016nhb,Lu:2016zhe, Esposito:2016itg,Albaladejo:2016eps,Ali:2016gdg}%
.

In the present article we continue our investigation of the exotic state $%
X_b $ by supposing that it can be considered as a meson molecule. In other
words, we assume that $X_b$ is a molecule-like state composed of $B^{+}$ and
$\overline{K}^{0}$ mesons. We are going to calculate its mass, decay
constant and width of the decay channel using the corresponding
molecule-type interpolating current and theoretical methods presented in
rather detailed form in Refs.\ \cite%
{Agaev:2016mjb,Agaev:2016ijz,Agaev:2016lkl,Agaev:2016dev}. Our aim is to
answer the question: has the exotic $X_{b}$ state observed by the D0
Collaboration the structure of a meson molecule or it is a tightly-bound
tetraquark state?

This work is organized in the following way. In Section \ref{sec:Mass} we
introduce the interpolating current employed in QCD sum rule calculations.
Here we derive the sum rules to evaluate the mass, decay constant and width
of the decay $X_{b}\to B_{s}\pi^{+}$. In Sect.\ \ref{sec:Num} we present
results of numerical calculations and compare them with experimental data
and theoretical predictions. This section contains also our conclusions.
Explicit expressions for the spectral density required for computation of
the mass and decay constant of the $X_b$ state are collected in Appendix A.


\section{The sum rules for the mass, current coupling and decay width}

\label{sec:Mass}

In this section we derive QCD sum rule expressions necessary to calculate
the mass, current coupling and width of the $X_{b}\to B_{s}\pi^{+}$ decay
employing a molecule-type interpolating current. The ways of calculations
have been considered in detailed form in our previous papers, therefore we
write down below only expressions that are new and differ from ones
presented there.

For calculation of the mass and current coupling we use the two-point QCD sum
rule and start from the correlation function
\begin{equation}
\Pi (p)=i\int d^{4}xe^{ip x}\langle 0|\mathcal{T}\{J^{X_{b}}(x)J^{X_{b}\dag
}(0)\}|0\rangle ,  \label{eq:CorrF1}
\end{equation}%
where $J^{X_{b}}(x)$ is the interpolating current with required quantum
numbers. We consider $X_{b}$ state as a particle with the quantum numbers $%
J^{P}=0^{+}$. In the meson molecule scheme the current $J^{X_{b}}(x)$ is
given by
\begin{equation}
J^{X_{b}}(x)=\left[ \overline{d}^{a}(x)\gamma _{5 }s^{a}(x)\right] \left[
\overline{b}^{b}(x)\gamma_{5}u^{b}(x)\right],  \label{eq:CDiq1}
\end{equation}
where $a$ and $b$ are color indices.

The standard procedures for deriving QCD sum rules include at first stage
computation of the correlation function in terms of the physical degrees of
freedom. The final expression, which one uses to get the relevant sum rule,
is the Borel transformed form of the function $\Pi ^{\mathrm{Phys}}(p)$. In
the case under consideration it is given as
\begin{equation}
\mathcal{B}_{p^{2}}\Pi ^{\mathrm{Phys}%
}(p)=m_{X_{b}}^{2}f_{X_{b}}^{2}e^{-m_{X_{b}}^{2}/M^{2}}+\ldots .
\label{eq:CorBor}
\end{equation}

The second step is to find the theoretical expression for the same function,
$\Pi ^{\mathrm{QCD}}(p)$, employing the quark-gluon degrees of freedom. To
this end, contracting the quark fields for the correlation function $\Pi ^{%
\mathrm{QCD}}(p)$ we find:
\begin{eqnarray}
&&\Pi ^{\mathrm{QCD}}(p)=i\int d^{4}xe^{ipx}\mathrm{Tr}\left[ \gamma
_{5}S_{s}^{aa^{\prime }}(x)\gamma _{5}S_{d}^{a^{\prime }a}(-x)\right]  \notag
\\
&&\times \mathrm{Tr}\left[ \gamma _{5}S_{u}^{bb^{\prime }}(x)\gamma
_{5}S_{b}^{b^{\prime }b}(-x)\right] ,  \label{eq:CorrF2}
\end{eqnarray}%
where $S_{q}^{ab}(x)$ and $S_{b}^{ab}(x)$ are the light ($q\equiv u,\ d\ $or
$s$) and $b$-quark propagators, respectively. We choose the light quark
propagator $S_{q}^{ab}(x)$ in the form%
\begin{eqnarray}
&&S_{q}^{ab}(x)=i\delta _{ab}\frac{\slashed x}{2\pi ^{2}x^{4}}-\delta _{ab}%
\frac{m_{q}}{4\pi ^{2}x^{2}}-\delta _{ab}\frac{\langle \overline{q}q\rangle
}{12}  \notag \\
&&+i\delta _{ab}\frac{\slashed xm_{q}\langle \overline{q}q\rangle }{48}%
-\delta _{ab}\frac{x^{2}}{192}\langle \overline{q}g\sigma Gq\rangle +i\delta
_{ab}\frac{x^{2}\slashed xm_{q}}{1152}\langle \overline{q}g\sigma Gq\rangle
\notag \\
&&-i\frac{gG_{ab}^{\alpha \beta }}{32\pi ^{2}x^{2}}\left[ \slashed x{\sigma
_{\alpha \beta }+\sigma _{\alpha \beta }}\slashed x\right] -i\delta _{ab}%
\frac{x^{2}\slashed xg^{2}\langle \overline{q}q\rangle ^{2}}{7776}  \notag \\
&&-\delta _{ab}\frac{x^{4}\langle \overline{q}q\rangle \langle
g^{2}G^2\rangle }{27648}+\ldots  \label{eq:qprop}
\end{eqnarray}%
For the $b$-quark propagator $S_{b}^{ab}(x)$ we employ the expression
from Ref.\ \cite{Reinders:1984sr}
\begin{eqnarray}
&&S_{b}^{ab}(x)=i\int \frac{d^{4}k}{(2\pi )^{4}}e^{-ikx} \Bigg \{ \frac{%
\delta _{ab}\left( {\slashed k}+m_{b}\right) }{k^{2}-m_{b}^{2}}  \notag \\
&&-\frac{gG_{ab}^{\alpha \beta }}{4}\frac{\sigma _{\alpha \beta }\left( {%
\slashed k}+m_{b}\right) +\left( {\slashed k}+m_{b}\right) \sigma _{\alpha
\beta }}{(k^{2}-m_{b}^{2})^{2}}  \notag \\
&&+\frac{g^{2}G^{2}}{12}\delta _{ab}m_{b}\frac{k^{2}+m_{b}{\slashed k}}{%
(k^{2}-m_{b}^{2})^{4}}+\frac{g^{3}G^{3}}{48}\delta _{ab}\frac{\left( {%
\slashed k}+m_{b}\right) }{(k^{2}-m_{b}^{2})^{6}}  \notag \\
&& \times \left[ {\slashed k}\left( k^{2}-3m_{b}^{2}\right) +2m_{b}\left(
2k^{2}-m_{b}^{2}\right) \right] \left( {\slashed k}+m_{b}\right) +\ldots %
\Bigg \}.  \notag \\
&& {}  \label{eq:Qprop}
\end{eqnarray}%
In Eqs.\ (\ref{eq:qprop}) and (\ref{eq:Qprop}) we use the notations
\begin{eqnarray}
&&G_{ab}^{\alpha \beta } = G_{A}^{\alpha \beta
}t_{ab}^{A},\,\,~~G^{2}=G_{\alpha \beta }^{A}G_{\alpha \beta }^{A},  \notag
\\
&&G^{3} =\,\,f^{ABC}G_{\mu \nu }^{A}G_{\nu \delta }^{B}G_{\delta \mu }^{C},
\end{eqnarray}%
where $a,\,b=1,2,3$ and $A,B,C=1,\,2\,\ldots 8$ are the color indices, and $%
t^{A}=\lambda ^{A}/2$ with $\lambda ^{A}$ being the Gell-Mann matrices. In
the nonperturbative terms the gluon field strength tensor $G_{\alpha \beta
}^{A}\equiv G_{\alpha \beta }^{A}(0)$ is fixed at $x=0.$

The correlation function $\Pi ^{\mathrm{QCD}}(p^{2})$ is given by a simple
dispersion integral
\begin{equation}
\Pi ^{\mathrm{QCD}}(p^{2})=\int_{(m_{b}+m_{s})^{2}}^{\infty }\frac{\rho ^{%
\mathrm{QCD}}(s)}{s-p^{2}}+...,  \label{CFQCD}
\end{equation}%
where $\rho ^{\mathrm{QCD}}(s)$ is the corresponding spectral density. We
have calculated the spectral density by including into analysis the quark,
gluon and mixed condensates up to eight dimensions. Our result for $\rho^{\mathrm{QCD}}(s)$ 
is moved to Appendix A.

Applying the Borel transformation on the variable $p^{2}$ to the invariant
amplitude $\Pi ^{\mathrm{QCD}}(p^{2})$, equating the obtained expression to $%
\mathcal{B}_{p^{2}}\Pi ^{\mathrm{Phys}}(p)$, and subtracting the continuum
contribution, we finally obtain the required sum rules. Thus, the mass of
the $X_{b}$ state can be evaluated from the sum rule
\begin{equation}
m_{X_{b}}^{2}=\frac{\int_{(m_{b}+m_{s})^{2}}^{s_{0}}dss \rho ^{\mathrm{QCD}%
}(s)e^{-s/M^{2}}}{\int_{(m_{b}+m_{s})^{2}}^{s_{0}}ds\rho ^{\mathrm{QCD}%
}(s)e^{-s/M^{2}}},  \label{eq:srmass}
\end{equation}%
whereas for the decay constant $f_{X_{b}}$ we employ the formula
\begin{equation}
f_{X_{b}}^{2}m_{X_{b}}^{2}e^{-m_{X_{b}}^{2}/M^{2}}=%
\int_{(m_{b}+m_{s})^{2}}^{s_{0}}ds\rho ^{\mathrm{QCD}}(s)e^{-s/M^{2}}.
\label{eq:srcoupling}
\end{equation}

In order to find the width of $X_b \to B_{s} \pi$ decay, we start from
calculation of the strong coupling $g_{X_bB_s \pi }$ using QCD sum rule on
the light-cone and soft-meson approximation. We consider the correlation
function
\begin{equation}
\Pi(p,q)=i\int d^{4}xe^{ipx}\langle \pi (q)|\mathcal{T}%
\{J^{B_{s}}(x)J^{X_{b}\dag }(0)\}|0\rangle ,  \label{eq:CorrF3}
\end{equation}
that allows us to get the sum rule for the coupling $g_{X_{b}B_s \pi }$.
Here the interpolating current $J^{B_s}(x)$ is defined in the form:
\begin{equation}
J^{B_s}(x)=\overline{b}_{l}(x)i\gamma _{5 }s_{l}(x).  \label{eq:Bcur}
\end{equation}

In the soft-meson limit $q=0$, for the Borel transformed form of the
correlation function $\Pi ^{\mathrm{Phys}}(p,q=0)$, we get (see, Refs.\ \cite%
{Agaev:2016ijz,Agaev:2016dev}),
\begin{eqnarray}
&&\Pi ^{\mathrm{Phys}}(M^{2})=\frac{%
f_{B_{s}}f_{X_{b}}m_{X_{b}}m_{B_{s}}^{2}g_{X_{b}B_{s}\pi }}{(m_{b}+m_{s})}%
m^{2}  \notag \\
&&\times \frac{1}{M^{2}}e^{-m^{2}/M^{2}}.
\end{eqnarray}%
where $m^2=(m_{X_b}^{2}+m_{B_{s}}^{2})/2$.

To proceed, we have to calculate $\Pi ^{\mathrm{QCD}}(p,q)$ in terms of the
quark-gluon degrees of freedom and find QCD side of the sum rule.
Contractions of $s$ and $b$-quark fields in Eq.\ (\ref{eq:CorrF3}) yield
\begin{eqnarray}
&&\Pi ^{\mathrm{QCD}}(p,q)=-\int d^{4}xe^{ipx}\left[
\gamma_{5}S_{b}^{bi}(-x){}\gamma _{5}\right.  \notag \\
&&\left. \times S_{s}^{ia}(x){}\gamma _{5}\right] _{\alpha \beta }\langle
\pi (q)|\overline{u}_{\alpha }^{b}(0)d_{\beta }^{a}(0)|0\rangle ,
\label{eq:CorrF6}
\end{eqnarray}%
where $\alpha $ and $\beta $ are the spinor indices. Now we use the
expansion
\begin{equation}
\overline{u}_{\alpha }^{b}d_{\beta }^{a}\rightarrow \frac{1}{4}\Gamma
_{\beta \alpha }^{j}\left( \overline{u}^{b}\Gamma ^{j}d^{a}\right) ,
\label{eq:MatEx}
\end{equation}%
where $\Gamma ^{j}$ is the full set of Dirac matrixes
\begin{equation*}
\Gamma ^{j}=\mathbf{1,\ }\gamma _{5},\ \gamma _{\lambda },\ i\gamma
_{5}\gamma _{\lambda },\ \sigma _{\lambda \rho }/\sqrt{2},
\end{equation*}%
and determine the required local matrix elements. For this purpose, we first
perform summation over the color indices. To clarify the computational
scheme, let us take a term $\sim \delta _{bi}$ from the b-quark propagator
and terms $\sim \delta _{ia}$ from the s-quark propagator considering by
this way, terms without gluon contributions. Then we get
\begin{equation}
\delta _{bi}\delta _{ia}\overline{u}_{\alpha }^{b}d_{\beta }^{a}=\delta _{ba}%
\frac{\delta _{ba}}{3}\overline{u}_{\alpha }d_{\beta }=\overline{u}_{\alpha
}d_{\beta }.
\end{equation}
Stated differently, after color summation in these terms we have to use the
replacement
\begin{equation}
\overline{u}_{\alpha }^{b}d_{\beta }^{a}\Rightarrow \overline{u}_{\alpha
}d_{\beta }.
\end{equation}
Now let us consider contributions $\sim G$. We may take a part $\sim \delta $
from one propagator and nonperturbative part from the another one. Then, as
an example, for the color structure of such term we find
\begin{equation}
\delta _{bi}gG_{ai}^{\varkappa \omega }\overline{u}_{\alpha }^{b}d_{\beta
}^{a}=gG_{ba}^{\varkappa \omega } \overline{u}_{\alpha}^{b}d^{a}_{\beta} =g
\overline{u}_{\alpha}G^{\varkappa \omega }d_{\beta}.
\end{equation}
Therefore, we get a rule
\begin{equation*}
\delta _{bi}gG_{ai}^{\varkappa \omega }\overline{u}_{\alpha }^{b}d_{\beta
}^{a}\Rightarrow g\left( \overline{u}G^{\varkappa \omega }d\right).
\end{equation*}
This rule allows us to insert into quark matrix elements the gluon field
strength tensor $G$ that effectively leads to three-particle components and
corresponding matrix elements of the pion: in the present work we neglect
terms $\sim G^{2}$ and  $\sim G^{3}$ . As is seen, in the case of the molecule current the
color summation is trivial. One needs only to remove color factors from the
propagators and use the prescriptions given above.

Omitting technical details, which can be found in Refs.\ \cite%
{Agaev:2016ijz,Agaev:2016dev}, we provide final expression for the spectral
density, which is given as a sum of the perturbative and nonperturbative
components
\begin{equation}
\rho _{\mathrm{c}}^{\mathrm{QCD}}(s)=\rho _{\mathrm{c}}^{\mathrm{pert.}%
}(s)+\rho _{\mathrm{c}}^{\mathrm{n.-p.}}(s).  \label{eq:SD}
\end{equation}%
where
\begin{equation}
\rho _{\mathrm{c}}^{\mathrm{pert.}}(s)=\frac{f_{\pi }\mu _{\pi }}{32\pi ^{2}}%
\left[ s-2m_{b}(m_{b}-m_{s})\right] \sqrt{1-\frac{4m_{b}^{2}}{s}},
\label{eq:SD1}
\end{equation}%
and
\begin{eqnarray}
&&\rho _{\mathrm{c}}^{\mathrm{n.-p.}}(s)=\frac{f_{\pi }\mu _{\pi }}{24}%
\langle \overline{s}s\rangle \left[ sm_{s}\delta
^{^{(1)}}(s-m_{b}^{2})-2m_{b}\delta (s-m_{b}^{2})\right]   \notag \\
&&+\frac{f_{\pi }\mu _{\pi }}{144}\langle \overline{s}g\sigma Gs\rangle
\left\{ 6(m_{b}-m_{s})\delta ^{^{(1)}}(s-m_{b}^{2})+3s(m_{b}-2m_{s})\right.
\notag \\
&&\left. \times \delta ^{(2)}(s-m_{b}^{2})-s^{2}m_{s}\delta
^{(3)}(s-m_{b}^{2})\right\} .  \label{eq:SD2}
\end{eqnarray}%
In Eq.\ (\ref{eq:SD2}) $\delta ^{(n)}(s-m_{b}^{2})=(d/ds)^{n}\delta
(s-m_{b}^{2})$ that appear when extracting the imaginary part of the pole
terms.

As is seen, in the soft limit, the spectral density depends only the
parameters $f_{\pi}$ and $\mu_{\pi}$ through the pion's local matrix element
\begin{equation}
\langle 0|\overline{d}(0)i\gamma _{5}u(0)|\pi (q)\rangle =f_{\pi }\mu _{\pi
},  \label{eq:MatE2}
\end{equation}%
where
\begin{equation}
\mu _{\pi }=\frac{m_{\pi }^{2}}{m_{u}+m_{d}}=-\frac{2\langle \overline{q}%
q\rangle }{f_{\pi }^{2}}.  \label{eq:PionEl}
\end{equation}

The continuum subtraction is performed in a standard manner after $\rho
^{h}(s)\rightarrow \rho_{c}^{\mathrm{QCD}}(s)$ replacement. Then, the final
sum rule to evaluate the strong coupling reads
\begin{eqnarray}
&&g_{X_{b}B_{s}\pi }=\frac{(m_{b}+m_{s})}{%
f_{B_{s}}f_{X_{b}}m_{X_{b}}m_{B_{s}}^{2}m^{2}}\left( 1-M^{2}\frac{d}{dM^{2}}%
\right) M^{2}  \notag \\
&&\times \int_{(m_{b}+m_{s})^{2}}^{s_{0}}dse^{(m^{2}-s)/M^{2}}\rho_{c}^{%
\mathrm{QCD}}(s).  \label{eq:SRules}
\end{eqnarray}

The width of the decay $X_{b}\rightarrow B_{s}^{0}\pi^{+}$ can be found
applying the standard methods. Our calculations give%
\begin{eqnarray}
&&\Gamma \left( X_{b}\rightarrow B_{s}^{0}\pi ^{+}\right) =\frac{%
g_{X_{b}B_{s}\pi }^{2}m_{B_{s}}^{2}}{24\pi }\lambda \left( m_{X_{b}},\
m_{B_{s}},m_{\pi }\right)  \notag \\
&&\times \left[ 1+\frac{\lambda ^{2}\left( m_{X_{b}},\ m_{B_{s}},m_{\pi
}\right) }{m_{B_{s}}^{2}}\right] ,  \label{eq:DW}
\end{eqnarray}%
where
\begin{equation*}
\lambda (a,\ b,\ c)=\frac{\sqrt{a^{4}+b^{4}+c^{4}-2\left(
a^{2}b^{2}+a^{2}c^{2}+b^{2}c^{2}\right) }}{2a}.
\end{equation*}%
Equations\ (\ref{eq:SRules}) and (\ref{eq:DW}) are final expressions that
will be used for numerical analysis of the decay channel $X_{b}\rightarrow
B_{s}^{0}\pi ^{+}$.


\section{Numerical results and conclusions}

\label{sec:Num}
The QCD sum rules derived above contain numerous parameters (see, Table \ref{tab:Param}),
which have to be fixed in accordance with the usual procedures.
Thus, for numerical computation of the $X_{b}$ state's mass and decay constant
we need values of the quark, gluon and mixed condensates. For  the various vacuum condensates
we use their well known values. In this range the gluon condensate  $\langle g^3G^3\rangle$ is relatively
new parameter, for which we employ the estimate given in Ref.\ \cite{Narison:2015nxh}.
The QCD sum rules contain also $b$ and $s$ quark masses, and depend on $B_{s}$ meson's mass and
decay constant $f_{B_s}$. In the present work, we choose the mass of the $b$ quark in the $\overline{MS}$ scheme at the
scale $\mu=m_b$, whereas the decay constant $f_{B_s}$ is borrowed from the lattice calculations
from Ref.\ \cite{Bazavov1:2011aa}.

\begin{figure}[h]
\includegraphics[width=8.2cm]{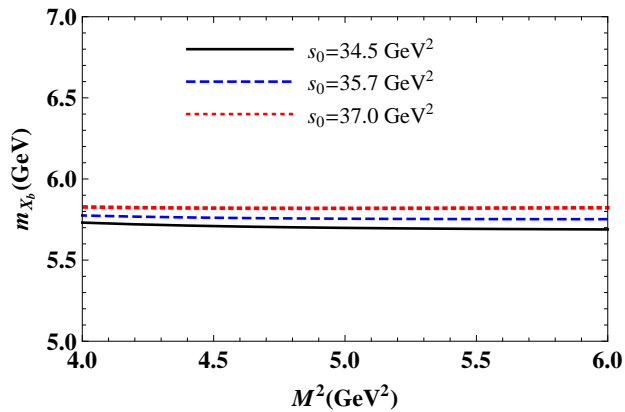}
\caption{The mass of the $X_{b}$ state versus the Borel parameter $M^{2}$.}
\label{fig:Mass}
\end{figure}

\begin{figure}[h]
\includegraphics[width=8.2cm]{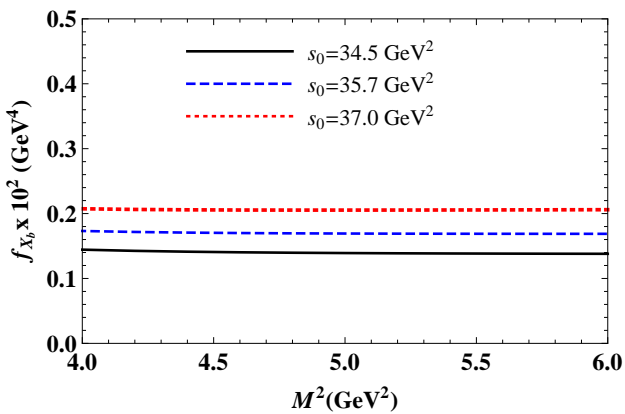}
\caption{The decay constant $f_{X_{b}}$ as a function the Borel parameter $%
M^{2}$.}
\label{fig:DecConst}
\end{figure}

The expressions evaluated within the QCD sum rule method additionally depend
on the continuum threshold and Borel parameters, i.e. on $s_{0}$ and $M^{2}$%
, respectively. One needs to fix some regions where physical quantities
under consideration are practically independent or demonstrate weak
dependence on them. To find the working window for the Borel parameter, we
require the convergence of the operator product expansion, as well as
suppression of the contributions arising from the higher resonances and
continuum, in other words exceeding of the pole contribution over the ones
coming from the higher dimensional condensates. As a result, for the mass
and current coupling calculations we find the range of $M^2$ as
\begin{equation}
4\ \mathrm{GeV}^{2}\leq M^{2}\leq 6\ \mathrm{GeV}^{2}.
\end{equation}%
The continuum threshold depends on the energy of the first excited state
with the same quantum numbers and structure as the particle under
consideration. In the case of the exotic states, it is difficult to
determine unambiguously this energy level, therefore we have to follow
standard recipes adopted in the sum rule computations and vary $s_{0}$
within the following region by considering the resonance of the D0
Collaboration as a ground state
\begin{equation}
34.5\,\,\mathrm{GeV}^{2}\leq s_{0}\leq 37\,\,\mathrm{GeV}^{2}.
\end{equation}%
Results of our numerical calculations of $m_{X_{b}}$ and $f_{X_{b}}$ are
depicted in Figs.\ \ref{fig:Mass} and \ref{fig:DecConst}. It is not
difficult to conclude that, in the chosen region the dependence of the mass and
decay constant on the parameter $M^{2}$ is insignificant. At the same time, obtained
predictions depend on the threshold parameter $s_0$, which
is the main source of the uncertainty of the sum rule computations. Variations of other
parameters within allowed limits give rise to errors, as well. The using in the leading order
sum rule expression the running quark  mass  $m_b(\mu)$, which in the lack of an information
on the next-to-leading order correction necessary to fix the renormalization scheme and
scale, also generates ambiguity. We treat the ambiguity arising because of the choice of the
renormalization scale $\mu$ as an additional source of the theoretical error, and include
its effect to the total error of the sum rule results presented below.

Hence, considering $X_{b}$ as the molecule state composed of the $B$ and $%
\overline{K}$ mesons, for its mass we obtain
\begin{equation}
m_{X_{b}}=(5757\pm 145)~\mathrm{MeV},
\end{equation}%
whereas the experimental value of the D0 Collaboration is equal to
\begin{equation}
m_{X_{b}}=5567.8\pm 2.9\mathrm{(stat)}_{-1.9}^{+0.9}\mathrm{(syst)}\,\mathrm{%
MeV}.
\end{equation}%
For the decay constant $f_{X_{b}}$ we get
\begin{equation}
f_{X_{b}}=(0.17\pm 0.04)\cdot 10^{-2}~\mathrm{GeV}^{4}.
\end{equation}%
The coupling $g_{X_{b}B_{s}\pi }$ extracted from the sum rule expression
Eq.\ (\ref{eq:SRules}) reads
\begin{equation}
g_{X_{b}B_{s}\pi }=(0.50\pm 0.14)~\mathrm{GeV}^{-1}.
\end{equation}%
As a result, for the width of the decay channel $\Gamma (X_{b}\rightarrow
B_{s}\pi )$ we find
\begin{equation}
\Gamma (X_{b}\rightarrow B_{s}\pi )=(33.6\pm 12.1)\ \mathrm{MeV},
\end{equation}%
when the experimental data give
\begin{equation}
\Gamma (X_{b}\rightarrow B_{s}\pi )=21.9\pm 6.4\mathrm{(stat)}_{-2.5}^{+5.0}%
\mathrm{(syst)}\,\mathrm{MeV}.
\end{equation}%
\begin{table}[tbp]
\begin{tabular}{|c|c|}
\hline\hline
Parameters & Values \\ \hline\hline
$m_{B_s}$ & $(5366.77\pm0.24)~\mathrm{MeV}$ \\
$f_{B_s}$ & $(242\pm 10)~\mathrm{MeV}$ \\
$m_{\pi}$ & $139.57 ~\mathrm{MeV}$ \\
$f_{\pi}$ & $0.131~\mathrm{GeV}$ \\
$m_{b}$ & $(4.18\pm0.03)~\mathrm{GeV}$ \\
$m_{s} $ & $(95 \pm 5)~\mathrm{MeV} $ \\
$\langle \bar{q}q \rangle $ & $(-0.24\pm 0.01)^3$ $\mathrm{GeV}^3$ \\
$\langle \bar{s}s \rangle $ & $0.8\ \langle \bar{q}q \rangle$ \\
$m_{0}^2 $ & $(0.8\pm0.1)$ $\mathrm{GeV}^2$ \\
$\langle\frac{\alpha_sG^2}{\pi}\rangle $ & $(0.012\pm0.004)$ $~\mathrm{GeV}%
^4 $ \\
$\langle g^3G^3\rangle $ & $(0.57\pm0.29)$ $~\mathrm{GeV}%
^6 $ \\
 \hline\hline
\end{tabular}%
\caption{Input parameters.}
\label{tab:Param}
\end{table}

As is seen, the values obtained for the mass and decay width of $X_b$ state
by treating it as a meson molecule state, overshoot corresponding
experimental data. This discrepancy is essential in the case of the mass,
which is slightly less than the mass threshold of the $B\overline{K}$ system
$\simeq 5778\ \mathrm{MeV}$, and differs significantly from the experimental
value of the D0 Collaboration. At the same time, the corresponding
predictions obtained in Ref.\ \cite{Agaev:2016mjb,Agaev:2016ijz} using the
diquark-antidiquark model for the $X_b$ state lead to a good agreement with
experimental data of the D0 Collaboration.

The QCD sum rule predictions for the mass and decay width extracted in the
present work by employing the molecule type interpolating current suffer
from the large uncertainties. But such errors are inherent in the sum rule
calculations, and are unavoidable part of the whole picture. The same
conclusion is valid for the diquark-antidiquark current, as well. These two
results have large overlap region making conclusive decision on the nature
of the $X_b$ state rather problematic. Nevertheless, our results for the
parameters of the $X_b$ state derived in the meson-molecule picture and
their comparison with ones evaluated by applying the diquark-antidiquark
model strengthen our confidence that if an exotic state $X_b$ with the
parameters of the D0 Collaboration exists, the
diquark-antidiquark form for its internal organization is more acceptable
than configuration of a molecule built of the $B$ and $\overline{K}$ mesons.

In the present work we have performed QCD sum rule analysis of the exotic $%
X_{b}$ state by treating it as a molecule composed of the $B$ and $\overline{%
K}$ mesons. We have calculated the mass $m_{X_{b}}$ and width of the decay $%
\Gamma (X_{b}\rightarrow B_{s}\pi )$, and compared our results with the
experimental data, as well as with predictions of the diquark-antidiquark
picture. From the present analysis we conclude that the molecule model for
the exotic $X_{b}$ state is less suitable than the diquark-antidiquark one
to explain its parameters measured by the D0 Collaboration. Further
investigations are required to clarify, first of all, the experimental
situation emerged because of the information of the LHCb Collaboration.
Theoretical studies have to be concentrated on preparing reliable models and
computational schemes to treat such complicated many-quark systems like the
exotic states.

\section*{ACKNOWLEDGEMENTS}

The work of S.~S.~A. was supported by the TUBITAK grant 2221-"Fellowship
Program For Visiting Scientists and Scientists on Sabbatical Leave". This
work was also supported in part by TUBITAK under the grant no: 115F183.

\appendix*

\section{A}

\renewcommand{\theequation}{\Alph{section}.\arabic{equation}}

\label{sec:App} In this appendix we have collected the results of our
calculations of the spectral density
\begin{equation}
\rho ^{\mathrm{QCD}}(s)=\rho ^{\mathrm{pert}}(s)+\sum_{k=3}^{8}\rho
_{k}(s),  \label{eq:A1}
\end{equation}%
used for evaluation of the $X_{b}$ meson mass $m_{X_{b}}$ and its decay
constant $f_{X_{b}}$ from the QCD sum rule. In Eq.\ (\ref{eq:A1}) by $\rho
_{k}(s)$ we denote the nonperturbative contributions to $\rho ^{\mathrm{QCD}%
}(s)$. In calculations we have neglected the masses of the $u$ and $d$
quarks and taken into account terms $\sim m_s$. The explicit expressions for
$\rho ^{\mathrm{pert}}(s)$ and $\rho _{k}(s)$ are presented below as
integrals over the Feynman parameter $z$.
\begin{widetext}
\begin{eqnarray}
&&\rho ^{\mathrm{pert}}(s)=\frac{1}{8192\pi ^{6}}\int\limits_{0}^{a}\frac{%
dzz^{4}}{(1-z)^{3}}\left[ m_{b}^{2}+s(z-1)\right] ^{3}\left[
m_{b}^{2}+3s(z-1)\right] ,  \notag \\
&&\rho _{\mathrm{3}}(s)=\frac{3}{256\pi ^{4}}\int\limits_{0}^{a}\frac{dzz^{2}%
}{(z-1)^{2}}\left[ m_{b}^{2}+s(z-1)\right] \Big \{ -2\langle \overline{d}%
d\rangle m_{s}(1-z)\left[ m_{b}^{2}+2s(z-1)\right] -m_{b}^{3}\langle
\overline{u}u\rangle +m_{b}^{2}m_{s}\langle \overline{s}s\rangle
(1-z)   \notag \\
&& -2m_{s}s\langle \overline{s}s\rangle (z-1)^{2}+m_{b}s\langle
\overline{u}u\rangle \ (1-z)\Big \} ,  \notag \\
&&\rho _{\mathrm{4}}(s)=\frac{1}{12288\pi ^{4}}\langle \alpha _{s}\frac{G^{2}%
}{\pi }\rangle \int\limits_{0}^{a}\frac{dzz^{2}}{(1-z)^{3}}\Big \{
2m_{b}^{4}(13z^{2}-30z+18)+3m_{b}^{2}s(6-5z)^{2}(z-1)+24s^{2}(z-1)^{3}(2z-3)%
\Big \} ,  \notag \\
&&\rho _{\mathrm{5}}(s)=\frac{m_{0}^{2}}{256\pi ^{4}}\int\limits_{0}^{a}%
\frac{dzz}{(1-z)}\Big \{ 3m_{s}\langle \overline{d}d\rangle (z-1)\left[
2m_{b}^{2}+3s(z-1)\right] -3m_{b}^{3}\langle \overline{u}u\rangle
-2m_{b}^{2}m_{s}\langle \overline{s}s\rangle (z-1)-3m_{b}\langle \overline{u}%
u\rangle s(z-1)  \notag \\
&& -3m_{s}s\langle \overline{s}s\rangle (z-1)^{2}\Big \} ,  \notag \\
&&\rho _{\mathrm{6}}(s)=\frac{1}{64\pi ^{4}}\int\limits_{0}^{a}dzz\Bigg \{
\frac{z^{4}}{5120\pi ^{2}(1-z)^{3}}\langle g^{3}G^{3}\rangle \left[
m_{b}^{2}(2z+3)+s(z-1)(5z-2)\right] -\frac{g^{2}}{27}\left[ \langle
\overline{u}u\rangle ^{2}+\langle \overline{d}d\rangle ^{2}+\langle
\overline{s}s\rangle ^{2}\right]    \notag \\
&& \times \left[ 2m_{b}^{2}+3s(z-1)\right] \Bigg \} , \\
&&\rho _{\mathrm{7}}(s)=-\frac{1}{768\pi ^{2}}\langle \alpha _{s}\frac{G^{2}}{%
\pi }\rangle \int\limits_{0}^{a}dz\frac{1}{(z-1)^{2}}\Big \{ 2m_{s}\langle
\overline{d}d\rangle (5z+1)(z-1)^{2}+m_{b}\langle \overline{u}u\rangle \left[
z(2z^{2}+7z-14)+7\right] -6m_{s}\langle \overline{s}s\rangle
z(z-1)^{2}\Big \} ,  \notag \\
&&\rho _{\mathrm{8}}(s)=-\frac{1}{16\pi ^{2}}\int\limits_{0}^{a}dz\left\{
\frac{m_{b}^{2}z^{2}}{384(z-1)^{2}}\langle \alpha _{s}\frac{G^{2}}{\pi }%
\rangle ^{2}\left[ s\delta ^{(1)}(s-m_{b}^{2}/(1-z))+2\delta
(s-m_{b}^{2}/(1-z))\right] \right.   \label{eq:ro2} \\
&&\left. +\frac{m_{0}^{2}m_{b}m_{s}}{16}\langle \overline{u}u\rangle \left[
12\langle \overline{d}d\rangle -5\langle \overline{s}s\rangle \right] \delta
(s-m_{b}^{2})-m_{0}^{2}\langle \overline{d}d\rangle \langle \overline{s}%
s\rangle (z-1)\right\} ,
\end{eqnarray}%
where $a=1-m_{b}^{2}/s$.
\end{widetext}


\begin{thebibliography}{99}

\bibitem{Belle:2003} S.-K.~Choi \textit{et al.} [Belle Collaboration],
Phys.\ Rev.\ Lett.\ \textbf{91}, 262001 (2003).


\bibitem{D0:2004} V.~M.~Abazov \textit{et al.} [D0 Collaboration], Phys.\
Rev.\ Lett.\ \textbf{93}, 162002 (2004).


\bibitem{CDF:2004} D.~Acosta \textit{et al.} [CDF II Collaboration] Phys.\
Rev.\ Lett.\ \textbf{93}, 072001 (2004).


\bibitem{Babar:2005} B.~Aubert \textit{et al.} [BaBar Collaboration], Phys.\
Rev.\ D\ \textbf{71}, 071103 (2005).


\bibitem{Swanson:2006st} E.~S.~Swanson,
Phys.\ Rept.\ \textbf{429}, 243 (2006). 


\bibitem{Klempt:2007cp} E.~Klempt and A.~Zaitsev,
Phys.\ Rept.\ \textbf{454}, 1 (2007). 


\bibitem{Godfrey:2008nc} S.~Godfrey and S.~L.~Olsen,
Ann.\ Rev.\ Nucl.\ Part.\ Sci.\ \textbf{58}, 51 (2008).


\bibitem{Voloshin:2007dx} M.~B.~Voloshin, 
Prog.\ Part.\ Nucl.\ Phys.\ \textbf{61}, 455 (2008).


\bibitem{Nielsen:2010} M.~Nielsen, F.~S.~Navarra, and S.~H.~Lee, Phys.\
Rep.\ \textbf{497}, 41 (2010).


\bibitem{Faccini:2012pj} R.~Faccini, A.~Pilloni and A.~D.~Polosa,
Mod.\ Phys.\ Lett.\ A \textbf{27}, 1230025 (2012).


\bibitem{Esposito:2014rxa} A.~Esposito, A.~L.~Guerrieri, F.~Piccinini,
A.~Pilloni and A.~D.~Polosa, 
Int.\ J.\ Mod.\ Phys.\ A \textbf{30}, 1530002 (2014).


\bibitem{Chen:2016} H.-X.~Chen, W.~Chen, X.~Liu, and S.-L.~Zhu, Arxiv:
1601.02092 [hep-ph], 2016.


\bibitem{D0:2016mwd} V.~M.~Abazov \textit{et al.} [D0 Collaboration],
arXiv:1602.07588 [hep-ex]. 


\bibitem{LHCb:2016} The LHCb Collaboration [LHCb Collaboration],
LHCb-CONF-2016-004, CERN-LHCb-CONF-2016-004.


\bibitem{Agaev:2016mjb} S.~S.~Agaev, K.~Azizi and H.~Sundu,
 Phys.\ Rev.\ D\ \textbf{93}, 074024 (2016).


\bibitem{Agaev:2016ijz} S.~S.~Agaev, K.~Azizi and H.~Sundu,
Phys.\ Rev.\ D\ \textbf{93}, 114007 (2016).


\bibitem{Agaev:2016lkl} S.~S.~Agaev, K.~Azizi and H.~Sundu,
 Phys.\ Rev.\ D\ \textbf{93}, 094006 (2016).


\bibitem{Wang:2016mee} Z.~G.~Wang,
arXiv:1602.08711 [hep-ph]. 


\bibitem{Wang:2016tsi} W.~Wang and R.~Zhu,
arXiv:1602.08806 [hep-ph]. 


\bibitem{Chen:2016mqt} W.~Chen, H.~X.~Chen, X.~Liu, T.~G.~Steele and
S.~L.~Zhu,
arXiv:1602.08916 [hep-ph]. 


\bibitem{Zanetti:2016wjn} C.~M.~Zanetti, M.~Nielsen and K.~P.~Khemchandani,
arXiv:1602.09041 [hep-ph]. 


\bibitem{Dias:2016dme} J.~M.~Dias, K.~P.~Khemchandani, A.~M.~Torres,
M.~Nielsen and C.~M.~Zanetti,
arXiv:1603.02249 [hep-ph]. 


\bibitem{Wang:2016wkj} Z.~G.~Wang,
arXiv:1603.02498 [hep-ph]. 


\bibitem{Xiao:2016mho} C.~J.~Xiao and D.~Y.~Chen,
arXiv:1603.00228 [hep-ph]. 


\bibitem{Liu:2016ogz} Y.~R.~Liu, X.~Liu and S.~L.~Zhu,
arXiv:1603.01131 [hep-ph]. 


\bibitem{He:2016yhd} X.~G.~He and P.~Ko,
arXiv:1603.02915 [hep-ph].


\bibitem{Jin:2016cpv} Y.~Jin and S.~Y.~Li,
arXiv:1603.03250 [hep-ph].


\bibitem{Stancu:2016sfd} F.~Stancu,
arXiv:1603.03322 [hep-ph].


\bibitem{Burns:2016gvy} T.~J.~Burns and E.~S.~Swanson,
arXiv:1603.04366 [hep-ph].


\bibitem{Tang:2016pcf} L.~Tang and C.~F.~Qiao,
arXiv:1603.04761 [hep-ph].


\bibitem{Guo:2016nhb} F.~K.~Guo, U.~G.~Mei?ner and B.~S.~Zou,
arXiv:1603.06316 [hep-ph].


\bibitem{Lu:2016zhe} Q.~F.~Lu and Y.~B.~Dong,
arXiv:1603.06417 [hep-ph].


\bibitem{Esposito:2016itg} A.~Esposito, A.~Pilloni and A.~D.~Polosa,
arXiv:1603.07667 [hep-ph].


\bibitem{Albaladejo:2016eps} M.~Albaladejo, J.~Nieves, E.~Oset, Z.~F.~Sun
and X.~Liu, 
arXiv:1603.09230 [hep-ph].


\bibitem{Ali:2016gdg} A.~Ali, L.~Maiani, A.~D.~Polosa and V.~Riquer,
arXiv:1604.01731 [hep-ph]. 

\bibitem{Agaev:2016dev} S.~S.~Agaev, K.~Azizi and H.~Sundu,
Phys.\ Rev.\ D\ \textbf{93}, 074002 (2016).

\bibitem{Reinders:1984sr} L.~J.~Reinders, H.~Rubinstein and S.~Yazaki,
Phys.\ Rept.\ \textbf{127}, 1 (1985).

\bibitem{Narison:2015nxh}
  S.~Narison,
  Nucl.\ Part.\ Phys.\ Proc.\  {\bf 270-272}, 143 (2016).


\bibitem{Bazavov1:2011aa} A.~Bazavov \textit{et al.} [Fermilab Lattice and
MILC Collaborations],
Phys.\ Rev.\ D \textbf{85}, 114506 (2012).

\end{thebibliography}
\end{document}